\documentclass[aps,twocolumn,amssymb,floatfix,amsmath]{revtex4}
\usepackage{graphicx}
\usepackage{dcolumn}
\usepackage{bm}

\begin{document}

\title{Coupling of electrons to the electromagnetic field in a localized basis}                              	
                                                                                     	
\author{Roland E. Allen\footnote{email allen@tamu.edu}}     	
                                                                                     	
\affiliation{Department of Physics, Texas A\&M University,                     	
College Station, Texas 77843}   
                                                                   	
\begin{abstract}                                                                     	
A simple formula is obtained for coupling electrons in a complex system 
to the electromagnetic field. It includes the effect of intra-atomic
excitations and nuclear motion, and can 
be applied in, e.g., first-principles-based simulations of the coupled dynamics 
of electrons and nuclei in materials and molecules responding to 
ultrashort laser pulses. Some additional aspects of
nonadiabatic dynamical simulations are also discussed, including the potential 
of ``reduced Ehrenfest'' simulations for treating problems where 
standard Ehrenfest simulations will fail.
\end{abstract}                  	
                                                                   	
\maketitle

It is now possible to perform first-principles simulations of the coupled
dynamics of electrons and nuclei with all nuclear coordinates 
included~\cite{joannopoulos,sankey,
TDDFT-surfaces,TDDFT_book}, rather than a subset of
nominal reaction coordinates. For very large systems, 
or when many trajectories are necessary, it is 
convenient to use a first-principles-\textit{based} 
scheme~\cite{frauenheim_1995,frauenheim_1998,
frauenheim_2002,torralva_2005}, with a valence-electron Hamiltonian and 
ion-ion repulsive potential derived from calculations using 
density-functional or other first-principles techniques. 
Here we are mainly concerned with the issue of how one can efficiently and
accurately couple electrons to the electromagnetic field in such an approach, 
where matrix elements of various operators between localized basis functions
(or ``atomic orbitals'') can be calculated from first principles, and 
then used in large-scale calculations for complex systems, such as 
materials and molecules, 
responding to applied fields, such as ultrashort laser 
pulses~\cite{mazur,Nature_fs,PRL_fs,zewail,graves,
dumitrica,torralva,dou,dou_2004,sauer_2007,sauer}.

Our starting point is, of course, the time-dependent Schr\"{o}dinger
equation
\begin{eqnarray}
&& i\hbar\frac{\partial}{\partial t}\psi\left(  \boldsymbol{x},t\right)
 = \hat{H}\,\psi\left(  \boldsymbol{x},t\right)  \label{SE} \\
\hspace {-1cm} \hat{H} &=& \frac{1}{2m}\left(  -i\hbar\boldsymbol{\nabla}-
\frac{q}{c}\boldsymbol{A}\left( \boldsymbol{x}, t \right) \right)^{2} 
+ U\quad ,\quad q=-e \label{hamil} \, .
\end{eqnarray}
Some time ago, Graf and Vogl~\cite{vogl} obtained a result, 
used in Refs.~\cite{graves,dumitrica,torralva,dou,dou_2004,sauer_2007,sauer}, 
which is the time-dependent version of the 
Peierls substitution: If $\boldsymbol{H}_{0}$ is the
Hamiltonian matrix in a localized basis with no applied field, 
\begin{eqnarray}
H_{0}\left(  \ell^{\prime}, \ell \right) = 
\int d^{3}x\;\phi_{a^{\prime}}^{\ast}\left( \boldsymbol{x} 
- \boldsymbol{X}^{\prime}\right) \hat{H}_{0}
\,\phi_{a}\left( \boldsymbol{x}-\boldsymbol{X}\right) \, ,
\end{eqnarray}
and $\tilde{\boldsymbol{H}}$ is the approximate Hamiltonian when there is an
applied field with vector potential $\boldsymbol{A}\left( \boldsymbol
{x},t\right)  $, then they are related by
\begin{eqnarray}
\tilde{H}\left(  \ell^{\prime},\ell\right) 
=H_{0}\left( \ell^{\prime},\ell \right)  e^{iq \bar{\boldsymbol{A}} 
\left( t \right) 
\cdot\left( \boldsymbol{X}^{\prime}-\boldsymbol{X}\right)  /\hbar c}
\label{vogl_hamil}
\end{eqnarray}
with
\begin{eqnarray}
\bar{\boldsymbol{A}} \left( t \right) = \left( \boldsymbol{A} 
\left( \boldsymbol{X}^{\prime} , t 
\right) + \boldsymbol{A} \left( \boldsymbol{X} , t \right)  \right) /2 .
\end{eqnarray}
Here $\ell$ labels a localized basis function centered on a nucleus
whose instantaneous position is $\boldsymbol{X} \left( \ell , t \right)$, 
and we adopt the convention of normally suppressing the indices $\ell$ and 
$\ell^{\prime}$, as well as the time $t$, by just writing $\boldsymbol{X}$
and $\boldsymbol{X}^{\prime}$. 
We will ignore any applied scalar potential $A_{0}$, any 
$\boldsymbol{\mu}_{B} \cdot \boldsymbol{B}$ spin interactions, and the coupling 
of ion cores or nuclei to the applied 
fields, since these effects can be easily included when necessary. 

With the prescription of (\ref{vogl_hamil}), one does not need any new parameters in 
a calculation that employs either a semiempirical~\cite{graves,dumitrica} or 
first-principles-based~\cite{torralva,dou,dou_2004,sauer_2007,sauer} 
Hamiltonian $\boldsymbol{H}_{0}$ whose elements are known as a function of
$\left(  \boldsymbol{X}-\boldsymbol{X}^{\prime}\right)  $. On the
other hand, this prescription is in one respect a rather crude approximation:
It omits intra-atomic excitations, and would therefore give no excitation at
all for isolated atoms. 

Here a more 
general version of the result of Ref.~\cite{vogl} will be obtained, in a form
which is almost equally convenient for large-scale applications, although it
does require new parameters -- namely dipole matrix elements
\begin{eqnarray}
\boldsymbol{\mu}_{0}\left(  \ell^{\prime} ,\ell\right) =  
q\int d^{3}x\;\phi_{a^{\prime}
}^{\ast}\left(  \boldsymbol{x}-\boldsymbol{X}^{\prime}\right)  \left(
\boldsymbol{x}-\boldsymbol{X}\right)  \phi_{a}\left(
\boldsymbol{x}-\boldsymbol{X}\right) \label{mu0}
\end{eqnarray}
and on-site ($\boldsymbol{X}^{\prime} = \boldsymbol{X}$) 
matrix elements of the momentum operator
\begin{eqnarray}
\boldsymbol{p}_{0}\left(  \ell^{\prime}
,\ell\right)  = \int d^{3}x\;\phi_{a^{\prime}}^{\ast}
\left(  \boldsymbol{x}-\boldsymbol{X}^{\prime} \right)  \left(
-i\hbar\boldsymbol{\nabla}\right)  \phi_{a}\left(  \boldsymbol
{x}-\boldsymbol{X}\right) \label{p0}
\end{eqnarray}
where $a$ labels an orbital centered on the nucleus whose instantaneous
position is $\boldsymbol{X}$. Recall that $\ell$ labels both nucleus
and orbital, so at a given instant in time
\begin{equation}
\ell \, \leftrightarrow \, \boldsymbol{X},a \, .
\end{equation}

One key step is to expand $\psi$ in terms of 
London orbitals, which we define to be any localized
basis functions $\tilde{\phi}_{a}$ that are related to 
field-independent basis functions $\phi_{a}$ by
\begin{equation}
\tilde{\phi}_{a}\left( \boldsymbol{x}-\boldsymbol
{X},t\right)  =e^{iq\boldsymbol{A}\left(  \boldsymbol{x},t\right)
\cdot\left(  \boldsymbol{x}-\boldsymbol{X}\right)  /\hbar c}
\phi_{a}\left(  \boldsymbol{x}-\boldsymbol{X}\right)
. \label{basis}
\end{equation}
Notice that $\tilde{\phi}_{a}\left(  \boldsymbol{x}-\boldsymbol
{X},t\right)  = \phi_{a}\left(  \boldsymbol{x}-\boldsymbol{X}\right)$
when $\boldsymbol{A}=0$, so that after the application
of a laser pulse, for example, the London orbitals return to being 
standard basis functions. 
The $\phi_{a}$ need not be a complete set, but should, of course, be 
a large enough set to model all physically relevant phenomena. The relatively 
weak time dependence of the nuclear positions $\boldsymbol{X}$ is ignored 
for the moment, but will be included below. The 
original Hamiltonian of (\ref{hamil}) can be rewritten as~\cite{vogl,peierls}
\begin{eqnarray}
\hat{H}&=&e^{iq \int \boldsymbol{A}\left( \boldsymbol{x},t \right)
\cdot d \boldsymbol{x}/\hbar c} \hat{H}_{0}\,e^{-iq \int \boldsymbol{A}\left(
\boldsymbol{x},t\right) \cdot d \boldsymbol{x}/\hbar c} \label{new_hamil}
\label{peierls} \\ 
\hat{H}_{0}&=&\hat{\boldsymbol{p}}^{2}/2m+U \quad , \quad \hat{\boldsymbol{p}}
= -i\hbar\boldsymbol{\nabla}
\end{eqnarray}
since (\ref{hamil}) and (\ref{new_hamil}) yield the same result when operating
on an arbitrary function, and are therefore the same operator. As will
be seen immediately below, there are no problems in interpreting the
integral of (\ref{peierls}) in the way that it is used here, since it 
is well-defined locally in evaluating each matrix element.

We now need the single approximation that
$\boldsymbol{A}\left( \boldsymbol{x},t \right) $ varies slowly with
respect to $\boldsymbol{x}$ over an atomic diameter or bond length, so
that
\begin{eqnarray}
\boldsymbol{A}\left( \boldsymbol{x},t \right) \approx 
\bar{\boldsymbol{A}}\left( t \right) \label{approx}
\end{eqnarray}
in the matrix elements which involve
$\phi_{a^{\prime}}\left( \boldsymbol{x}-\boldsymbol{X}^{\prime} \right)$ 
and $\phi_{a}\left( \boldsymbol{x}-\boldsymbol{X} \right)$.
(The wavelength is thus assumed to be large compared to $1$ \AA.) 
When
(\ref{new_hamil}) and
\begin{equation}
\psi\left(  \boldsymbol{x},t\right)  =\sum_{\ell}
\psi\left(  \ell,t\right)  \,\tilde{\phi}_{a
}\left(  \boldsymbol{x}-\boldsymbol{X},t\right)  \label{psi}
\end{equation}
are substituted into (\ref{SE}), and the resulting equation is subjected to
$\int d^{3}x\;\tilde{\phi}_{a^{\prime}}^{\ast}\left(  \boldsymbol
{x}-\boldsymbol{X}^{\prime},t\right)  $, we then obtain
\begin{equation}
\sum_{\ell} S \left(  \ell^{\prime},
\ell\right)  i\hbar\frac{\partial \psi\left(  \ell,t\right)}{\partial t}
=\sum_{\ell} H\left(  \ell^{\prime},
\ell\right)  \psi\left(  \ell,t \right)  \label{new_SE_1}
\end{equation}
where
\begin{eqnarray}
S\left(  \ell^{\prime},\ell\right)   
&=& S_{0}\left(  \ell^{\prime} ,\ell\right) 
e^{iq \bar{\boldsymbol{A}} \left( t \right) \cdot 
\left( \boldsymbol{X}^{\prime} - \boldsymbol{X} \right) /\hbar c} \\
S_{0}\left(  \ell^{\prime},\ell \right)   
&=& \int d^{3}x\;\phi_{a^{\prime}}^{\ast}\left(
\boldsymbol{x}-\boldsymbol{X}^{\prime}\right)  \phi_{a}\left(
\boldsymbol{x}-\boldsymbol{X}\right) \\
H\left( \ell^{\prime},\ell\right) &=& \tilde{H}\left( \ell^{\prime},\ell
\right) - \bar{\boldsymbol{E}} \left( t \right)
\cdot \boldsymbol{\mu}\left(\ell^{\prime},\ell \right) \label{full_H} \\
\boldsymbol{\mu}\left( \ell^{\prime},\ell\right) &=&  
\boldsymbol{\mu}_{0} \left(\ell^{\prime},\ell\right)
e^{iq \bar{\boldsymbol{A}} \left( t \right) \cdot 
\left( \boldsymbol{X}^{\prime} - \boldsymbol{X} \right)  /\hbar c} \label{mu}
\end{eqnarray}
and
\begin{equation}
\bar{\boldsymbol{E}} \left( t \right) =
-\frac{1}{c}\frac{\partial\bar{\boldsymbol{A}}\left( t \right)}
{\partial t} \label{E}
\end{equation}
is the electric field. In matrix form, (\ref{new_SE_1}) is
\begin{equation}
i\hbar\frac{\partial}{\partial t} \boldsymbol{\psi}\left(  t\right)
=\boldsymbol{S}^{-1}\cdot\boldsymbol{H}\cdot
\boldsymbol{\psi}\left(  t\right) \label{matrix_SE} .
\end{equation}
If there are $N_{e}$ electronic basis functions, then $\boldsymbol{\psi}$
is an $N_{e}$-dimensional vector, whereas $\boldsymbol{x}$,
$\boldsymbol{A}$, $\boldsymbol{\mu}$, etc. are 3-dimensional vectors. 
The dipole matrix elements can in principle be obtained in \textit{ab initio}
calculations like those used to obtain, e.g., the Hamiltonian matrix elements
$H_{0} \left(  \ell^{\prime},\ell\right)  $
\cite{frauenheim_1995,frauenheim_1998,frauenheim_2002,
torralva_2005}. Alternatively, one might make the approximation of 
including only the terms with single-atom dipole matrix elements, 
$\boldsymbol{\mu}_{0}\left( \boldsymbol{X}a^{\prime},\boldsymbol
{X}a\right)$, 
and then take these from either atomic calculations or experiment.

We now return to the time dependence of the nuclear positions 
$\boldsymbol{X}$. With (\ref{psi}) rewritten as 
\begin{eqnarray}
\psi\left(  \boldsymbol{x},t\right) &=&  
\sum_{\ell } \tilde{\psi}\left(  \ell,t\right)  \,\phi_{a
}\left(  \boldsymbol{x}-\boldsymbol{X} \right) \\ 
\tilde{\psi}\left( \ell,t \right)  &=& \psi \left( \ell,t\right) 
e^{iq\boldsymbol{A}\left(  \boldsymbol{x},t\right)
\cdot\left(  \boldsymbol{x}-\boldsymbol{X}\right)  /\hbar c}
\end{eqnarray}
we have~\cite{todorov} 
\begin{eqnarray}
\frac{\partial\psi}{\partial t} \hspace{-2.5pt} = \hspace{-2.5pt}
\sum_{\ell} \hspace{-2.5pt} \left[
\frac{\partial\tilde{\psi}\left(  \ell\right)
}{\partial t}\,\phi_{a}\left(  \boldsymbol{x}-\boldsymbol
{X}\right)  \hspace{-2.5pt} + \hspace{-2.0pt} \tilde{\psi}\left(  \ell\right)
\frac{\partial\phi_{a}\left(  \boldsymbol{x}-\boldsymbol
{X}\right)  }{\partial\boldsymbol{X}}\cdot \dot{\boldsymbol{X}}\right] 
\hspace{-3.5pt} .
\nonumber
\end{eqnarray}

In order to treat the second term above, we assume (as indicated by
the notation) that the basis functions depend only on 
$\left( \boldsymbol{x}-\boldsymbol{X} \right)$, so that
\begin{eqnarray}
\frac{\partial\phi_{a}\left( \boldsymbol{x}-\boldsymbol{X}\right)  }
{\partial\boldsymbol{X}} &=& -\frac{\partial\phi_{a}\left(
\boldsymbol{x}-\boldsymbol{X}\right)  }{\partial\left(
\boldsymbol{x}-\boldsymbol{X}\right)  } \\
&=& -\boldsymbol{\nabla}\phi_{a}\left(  \boldsymbol
{x}-\boldsymbol{X}\right)  .
\end{eqnarray}
There is an additional correction involving $\dot{\boldsymbol{X}}$
which arises from
\begin{eqnarray}
\frac{\partial\tilde{\psi}\left( \ell \right)}{\partial t} \hspace{-3.5pt}&=& 
\hspace{-3.5pt} e^{iq\boldsymbol{A}\left( \boldsymbol{x}\right)\cdot
\left( \boldsymbol{x}-\boldsymbol{X}\right)/\hbar c}
\Bigg[ \frac{\partial{\psi}\left( \ell \right)}{\partial t} \nonumber \\
\hspace{-3pt} &+& \hspace{-3pt} \psi 
\left( \ell \right) \left( \frac{iq}{\hbar c} \right) 
\left( \frac{\partial\boldsymbol{A}\left( \boldsymbol{x}\right)}{\partial t} 
\cdot \left( \boldsymbol{x}-\boldsymbol{X}\right) \hspace{-1.6pt} - 
\hspace{-1.6pt} \boldsymbol{A}\left( \boldsymbol{x}\right) 
\cdot \dot{\boldsymbol{X}} \right) \Bigg] . \nonumber 
\end{eqnarray}
It follows that (\ref{full_H}) is modified to
\begin{eqnarray}
H\left(  \ell^{\prime},\ell\right)    
 &=& H_{0}\left(  \ell^{\prime} ,\ell\right)  
e^{iq \bar{\boldsymbol{A}} \cdot \left( \boldsymbol{X}^{\prime} 
- \boldsymbol{X} \right)  /\hbar c}  \nonumber \\
 && - \bar{\boldsymbol{E}} \left( t \right)
\cdot\boldsymbol{\mu}\left( \ell^{\prime},
\ell\right)  - \dot{\boldsymbol{X}}\cdot
\boldsymbol{P} \left(  \ell^{\prime},\ell\right) \label{new_full_H}
\end{eqnarray}
where
\begin{eqnarray}
\boldsymbol{P}\left( \ell^{\prime},\ell \right) &=& 
\boldsymbol{p}\left( \ell^{\prime},\ell\right)
+ \left( q/c \right) \bar{\boldsymbol{A}} \;
S\left( \ell^{\prime},\ell\right) \\ 
\boldsymbol{p}\left(  \ell^{\prime}
,\ell\right) &=& \boldsymbol{p}_{0}\left( \ell^{\prime},\ell\right)
e^{iq \bar{\boldsymbol{A}} \cdot \left( \boldsymbol{X}^{\prime} 
- \boldsymbol{X} \right)  /\hbar c}   \label{p}
\end{eqnarray}
so another set of parameters is needed to treat the time dependence of the
basis functions that arises from nuclear motion -- namely, the matrix elements
of the momentum operator $-i\hbar \boldsymbol{\nabla}$. 

However, when $\boldsymbol{X}^{\prime }\neq \boldsymbol{X}$, there is a more
convenient way of writing $\boldsymbol{p}_{0}\left( \ell ^{\prime },\ell
\right) $:
\begin{eqnarray}
\boldsymbol{p}_{0}\left( \ell ^{\prime },\ell \right)  &=&i\hbar \int d^{3}x
\, \phi_{a ^{\prime }}^{\ast }\left( \boldsymbol{x}-\boldsymbol{X}^{\prime
}\right) \frac{\partial \phi _{a }\left( \boldsymbol{x}-\boldsymbol{X}
\right) }{\partial \boldsymbol{X}} \\
&=& i\hbar \frac{\partial }{\partial \boldsymbol{X}}S_{0}\left( \ell ^{\prime},
\ell \right) \quad \text{if} \quad \boldsymbol{X}^{\prime }\neq 
\boldsymbol{X}. \label{p0_off}
\end{eqnarray}

Furthermore, in the usual case of basis functions 
(``atomic orbitals'') which are
either even or odd under inversion through the nucleus, the fact that 
$\left( \boldsymbol{x}-\boldsymbol{X}\right)$ and $\boldsymbol{\nabla}=
\partial/\partial \left( \boldsymbol{x}-\boldsymbol{X}\right)$ 
are odd under inversion (with $\boldsymbol{X}$ here taken to be fixed)
implies that 
\begin{eqnarray}
\boldsymbol{\mu}_{0}\left(  \ell ,\ell\right) =  
\boldsymbol{p}_{0}\left(  \ell ,\ell\right) =  0 . \label{on}
\end{eqnarray}

Notice that (\ref{new_full_H}) respects gauge invariance: If 
\begin{eqnarray}
\bar{\boldsymbol{A}}\left( t \right) \rightarrow 
\bar{\boldsymbol{A}}^{\prime}\left( t \right) = 
\bar{\boldsymbol{A}}\left( t \right)  
+ \Delta \bar{\boldsymbol{A}}
\end{eqnarray}
where $\Delta \bar{\boldsymbol{A}}$ is independent of $t$, 
then (\ref{new_SE_1}) still holds with 
\begin{eqnarray}
\psi\left(  \ell,t \right) \rightarrow \psi^{\prime}\left( \ell,t \right)
=  e^{iq \Delta \bar{\boldsymbol{A}} \cdot \boldsymbol{X} /\hbar c} 
\psi\left(  \ell,t \right) .
\end{eqnarray}
This is the discrete version of 
\begin{eqnarray}
\boldsymbol{A} \left( \boldsymbol{x},t \right) &\rightarrow &
\boldsymbol{A}^{\prime} \left( \boldsymbol{x},t \right)  = 
\boldsymbol{A} \left( \boldsymbol{x},t \right) 
+ \nabla \Lambda \left( \boldsymbol{x} \right) \\
\psi \left( \boldsymbol{x},t \right) &\rightarrow &
\psi^{\prime} \left( \boldsymbol{x},t \right)  = 
e^{iq \Lambda \left( \boldsymbol{x} \right) /\hbar c} 
\psi \left( \boldsymbol{x},t \right) \, .
\end{eqnarray}
If $\Delta \bar{\boldsymbol{A}}$ is a function of $t$, gauge invariance
again holds, but with the scalar potential included.

Equation (\ref{new_full_H}) is the central result of the present note. 
This effective Hamiltonian is not manifestly Hermitian, but it still
conserves probability and preserves the Pauli principle, since a
straightforward calculation using (\ref{new_full_H}) in (\ref{new_SE_1})
gives
\begin{eqnarray}
i \hbar \; \partial \left( \boldsymbol{\psi}_{n^{\prime}}^{\dagger} \cdot 
\boldsymbol{S}  \cdot \boldsymbol{\psi}_{n} \right) / \partial t = 0
\end{eqnarray}
where $n$ labels a time-dependent one-electron state.
This result also follows from the original  Schr\"{o}dinger equation
(\ref{SE}) and the expansion (\ref{psi}), since
\begin{eqnarray}
\int d^{3}x\; \psi_{n^{\prime}}^{\ast} \left( \boldsymbol{x},t\right)
\psi_{n} \left( \boldsymbol{x},t\right) = 
\boldsymbol{\psi}_{n^{\prime}}^{\dagger}  
\left( t \right) \hspace{-2pt} \cdot \hspace{-2pt}
\boldsymbol{S} \left( t \right) \hspace{-2pt} \cdot \hspace{-1pt}
\boldsymbol{\psi}_{n} \left( t \right) \, ,
\end{eqnarray}
but it is reassuring that our approximation (\ref{approx}) preserves
orthonormality of the time-dependent states.

For slowly moving nuclei the last term in (\ref{new_full_H}) 
is not important. (It may be worth mentioning in this context 
that the direct coupling of 
the nuclei to the field is not considered here, since it can be
treated separately.) In an earlier paper~\cite{dumitrica_2004} 
we argued that the nuclear motion can be
approximately treated as a ``nuclear velocity
field'' analogous to the radiation field, and in this spirit
we obtained (as a crude approximation) a generalized Peierls substitution:
\begin{eqnarray}
\hspace{-0.6 cm} H_{eff}\left(  \ell^{\prime}, \ell \right) 
&=& e^{\frac{i}{\hbar}\left[  \frac{q}{c}
\boldsymbol{A}\left(  \boldsymbol{X}^{\prime}\right)  +m 
\dot {\boldsymbol{X}}^{\prime}\right]  \cdot\boldsymbol{X}
^{\prime} } \nonumber \\
& \times & H_{0}\left(  \ell^{\prime}
,\ell\right)  e^{-\frac{i}{\hbar}\left[  \frac
{q}{c}\boldsymbol{A}\left(  \boldsymbol{X}\right)  +m \dot
{\boldsymbol{X}}\right]  \cdot \boldsymbol{X} }.
\end{eqnarray}
We also used this modified Hamiltonian in calculations for organic molecules
responding to femtosecond-scale laser pulses of moderately strong intensity 
($\sim10^{12}$ W/cm$^{2}$), and found that the $\dot {\boldsymbol{X}}$ 
terms made very little difference in the final results. 
On the other hand, the 2-center momentum matrix elements can be 
obtained from  (\ref{p0_off}), and the nonzero 1-center matrix elements from 
either atomic calculations or experiment, so it is certainly feasible 
to include the last term in (\ref{new_full_H}).
Notice that this term is different from 
the Pulay correction~\cite{joannopoulos}, which also results 
from the fact that the basis functions follow the nuclei, but occurs in the 
equation of motion for the nuclei rather than the time-dependent 
Schr\"{o}dinger equation for the electrons. 
In the kind of approach considered here 
there is no Pulay correction, because the Hamiltonian matrix elements
are supposed to have a position dependence that includes the movement 
of the basis functions.

In this context, it is worth noting that the 
``Ehrenfest dynamics''~\cite{sakurai,allen_1994} 
of, e.g., time-dependent density-functional theory (TDDFT) and the 
density-functional-based calculations of 
Refs.~\cite{torralva,dou,dou_2004,sauer_2007,sauer}, 
can be substantially
improved in molecular calculations via a trivially different procedure 
that might be called ``reduced Ehrenfest dynamics'' and which is
similar in spirit to the surface hopping methods of Tully and 
others~\cite{tully,granucci}. Let 
us first recall some well-known results: The total wavefunction 
for a system of nuclei, with coordinates 
$X_{n}$, and electrons, with coordinates $x_{e}$, can be represented by the
Born-Oppenheimer expansion
\begin{equation}
\Psi ^{tot}\left(X_{n}, x_{e}, t\right) =\sum_{i}\Phi _{i}
\left(X_{n},t\right) \,\Psi _{i}\left( x_{e},X_{n}\right)
\label{BO_exp}.
\end{equation} 
The basis functions $\Psi _{i}$ are the 
electronic eigenstates at fixed $X_{n}$, with the electron-nuclei and
nuclei-nuclei interactions included in the electronic Hamiltonian $H_{e}$: 
\begin{equation}
H_{e}\left( X_{n}\right) \Psi _{i}\left( x_{e},X_{n}\right) 
=E_{i}\left( X_{n}\right) \Psi_{i} \left( x_{e},X_{n}\right) \, .
\end{equation}
Substitution into the Schr\"{o}dinger equation
\begin{eqnarray}
i\hbar \, \partial \Psi^{tot} / \partial t= \mathcal{H}\Psi
^{tot}\quad , \quad \mathcal{H} = T_{n} + H_{e} \, ,
\end{eqnarray}
where $T_{n}$ is the nuclear kinetic energy operator, gives an equation of
the form~\cite{conical,beyond}
\begin{eqnarray}
i\hbar \frac{\partial }{\partial t}\Phi _{i}  &=&\,
\left( T_{n } +E_{i} \right) \Phi _{i}
-\sum_{j}\Lambda _{ij}\Phi _{j} \label{coupled} \\ 
\Lambda _{ij} &=&\frac{\hbar^{2}}{2M_{n}}\left( 2\boldsymbol{F}_{ij}\cdot 
\boldsymbol{\nabla }_{n}+G_{ij}\right)  \\
\boldsymbol{F}_{ij} &=&\left\langle i\left\vert \boldsymbol{\nabla }_{n}
\right\vert j\right\rangle \quad ,\quad G_{ij}=\left\langle i\left\vert
\boldsymbol{\nabla}_{n} ^{2}\right\vert j\right\rangle 
\end{eqnarray}
where $M_{n}$ is a representative nuclear mass and $\boldsymbol{\nabla }_{n}$ 
involves all the appropriately rescaled nuclear coordinates.
If there are $N_{n}$ relevant nuclear coordinates, then 
$\boldsymbol{\nabla}_{n}$ and $\boldsymbol{F}$ are $N_{n}$-dimensional vectors. 
Also, quantities in the last line are matrix elements defined in terms of 
$\Psi _{i} ^{\dag}$ and $\Psi _{j}$ in the usual way. If the
components $\Phi _{i}$ are assembled into a vector $\Phi $, 
(\ref{coupled}) can be 
written in a form which resembles a nonabelian gauge theory~\cite{Zygelman}:
\begin{equation}
i\hbar \frac{\partial }{\partial t}\Phi =\,\left[ -
\frac{\hbar^{2}}{2M_{n}}\left( \boldsymbol{\nabla }_{n}+\boldsymbol{F}\right)^{2} 
+ \boldsymbol{E} \right] \cdot \Phi \label{nuclear}
\end{equation}
where $\boldsymbol{E}$ is the diagonal matrix with elements $E_{i}$. Finally, it
can be shown that~\cite{epstein}
\begin{equation}
\boldsymbol{F}_{ij}=\frac{\left\langle i\left\vert \boldsymbol
{\nabla }_{n} H_{e}\right\vert j\right\rangle }{E_{j}-E_{i}}
\quad , \quad E_{i} \neq E_{j} \; .
\end{equation}
This last equation implies that each term in the Born-Oppenheimer expansion
should evolve nearly independently if it is sufficiently distant in 
energy from all the other terms: If
\begin{eqnarray}
\left | E_{i}-E_{j} \right | \gg \left | \left\langle  
i\left\vert \boldsymbol{\nabla }_{n}H_{e}\right\vert j\right\rangle 
\right | \left( \hbar / P_{i} \right) \label{gg} 
\end{eqnarray}
where $P_{i} = \left( 2 M _{n} E_{i} \right)^{1/2}$, then
\begin{eqnarray}
i\hbar \, \partial \Phi _{i} / \partial t \approx
\left( T_{n} + E _{i} \right) \Phi _{i} \, . \label{BO_approx} 
\end{eqnarray}
This is the time-dependent Born-Oppenheimer or adiabatic approximation.

On the other hand, whenever nuclear motion causes two Born-Oppenheimer 
``potential energy surfaces'' to approach 
each other, so that 
\begin{eqnarray}
\left | E_{i}-E_{j} \right | \lesssim \left | \left\langle i
\left\vert \boldsymbol{\nabla }_{n}H_{e}\right\vert j\right\rangle
\right | \left( \hbar / P_{i} \right) \label{close}
\end{eqnarray}
there is a nonadiabatic interaction~\cite{conical,beyond,cederbaum,martinez},
and a Born-Oppenheimer simulation  based on (\ref{BO_approx}) is
invalid.

The results of Refs.~\cite{torralva,dou,dou_2004,sauer_2007,sauer} have
provided a clear demonstration of the following features of simulations
based on Ehrenfest dynamics:

(1) Electronic transitions are automatically observed at the
points of closest approach where ($\ref{close}$) holds, with 
energy released to molecular 
vibrations. These points are, of course, avoided crossings 
near the conical intersections in configuration space predicted by
Teller~\cite{conical,beyond,teller}.

(2) These transitions occur rapidly, over a time interval of $\sim 1$ 
femtosecond, during which the nuclei do not move appreciably.

Ehrenfest simulations are based on the equation of motion for the 
Heisenberg operator $\hat{X}\left( t \right) $ representing any nuclear
coordinate~\cite{sakurai,allen_1994}:
\begin{eqnarray}
M d^{2} \hat{X} / dt^{2} = -\partial {\cal H} / \partial \hat{X} \, .
\end{eqnarray}
Here $M$ is the corresponding nuclear mass and 
$\cal{H}$ is the Hamiltonian of the 
system. In a standard Ehrenfest simulation, the expectation value is
taken over the full state of the system, including excitations (e.g. by a
laser pulse) and de-excitations (e.g. by nuclear motion near conical 
intersections):
\begin{eqnarray}
M \frac{d^{2} \langle \hat{X} \rangle }{dt^{2}} = 
- \Bigg \langle \frac{\partial {\cal H} \left( \hat{X} \right)} 
{\partial \hat{X}} \Bigg \rangle \approx 
- \frac{\partial {\cal H} \left( \langle \hat{X} \rangle \right)} 
{\partial \langle \hat{X} \rangle} \, \label{classical} .
\end{eqnarray}

There are clearly two weaknesses with this approach: First, the
equality on the left represents an average over all the terms in the 
expansion (\ref{BO_exp}), with each term representing a different
nuclear trajectory. Second, the approximation on the right is totally 
invalid if these trajectories are very different.

Suppose, however, that the standard procedure for an Ehrenfest
simulation is replaced by a trivially different procedure in which the 
state of the system is collapsed to a single 
Born-Oppenheimer term immediately after an excitation or de-excitation 
event. Then (\ref{BO_approx}) implies that it will essentially remain 
in this single adiabatically evolving state until the next such event. 
For this reduced electronic state, the nuclei will ordinarily follow 
a single trajectory, except for quantum fluctuations of order 
$\Big \langle \left( \hat{X} - \langle \hat{X} \rangle \right)^{2} \Big \rangle$
~\cite{dpa}. It is still possible for nuclear wavepackets to diverge on a
single potential energy surface, but one does not expect this to be 
a common occurrence for processes in which the most relevant nuclei
are reasonably heavy. 

For simplicity, first consider a very short laser pulse (e.g. $\sim 1-5$ 
femtoseconds in duration) applied to a molecule. The procedure for a 
reduced Ehrenfest simulation is as follows: Start with a single 
electronic eigenstate (e.g. the ground 
state) and initially perform an Ehrenfest simulation in the usual way. 
Immediately following the pulse, the molecule will be in a
superposition of electronic eigenstates:
\begin{equation}
\Psi_{e} \left( t \right) =\sum_{i} c_{i} \, \Psi _{i} \, . 
\label{superposition}
\end{equation} 
At this point one collapses $\Psi_{e}$ to a single eigenstate $\Psi_{i}$ 
and continues the simulation, with $\langle \hat{X} \rangle$ now 
interpreted as the expectation value for this single resulting
time-dependent state, until another 
significant excitation or de-excitation is observed,
after which there is again a further reduction to a single electronic
eigenstate.

There are potentially a substantial number of branches to be followed 
during this sort of simulation, correponding to the various states in 
the superposition (\ref{superposition}) after an excitation or 
de-excitation event. The goal, however, is to
understand the most relevant processes, and there will ordinarily be
physical motivations for selecting the most interesting branches.
Similarly, there will be many branches emerging \textit{during} an excitation
process whose duration is long enough for the nuclei to move appreciably 
before it is completed (e.g., a femtosecond-scale 
laser pulse whose duration is still $\gg 1$ femtosecond), and a choice
among the branches again has to be based on physical considerations.

For a molecule subjected to high-frequency or high-intensity radiation, 
the branches include 
ionized states. The one-electron matrix element between an orbital $\ell$ 
and an ionized state with momentum $\boldsymbol{p}$ is 
\begin{eqnarray}
H_{\ell \, \boldsymbol{p}} = \frac{e}{mc} 
\boldsymbol{A} \left( \boldsymbol{X}, t \right)
\cdot \langle \ell \left| \, \hat {\boldsymbol{p}} \, 
\right| \boldsymbol{p} \rangle \, .
\end{eqnarray}
For a crude description of ionization, one might add a model 
orbital $\phi_{0}$ to the basis, with
\begin{eqnarray}
H_{0 \, \ell}= \alpha_{\ell} \, \frac{e}{mc} \left| \boldsymbol{A} 
\left( \boldsymbol{X}, t \right) \right| p_{0} \quad , 
\quad H_{\ell \, 0}= 0
\end{eqnarray}
where $p_{0} \sim \hbar /a_{0}$, $a_{0}$ is the Bohr radius, and 
$\alpha_{\ell}$ is an adjustable dimensionless parameter. This non-Hermitian
Hamiltonian removes amplitude from the orbital $\ell$ at each time
step and does not return it, so it crudely models excitation to a
localized wavepacket with the electron ultimately escaping the system.
An appreciable probability for a given ionized state then provides
motivation for following that branch in a reduced Ehrenfest simulation.
Notice that an accurate treatment of ionization is not necessary if the 
only issue is whether an ionized state is important enough to
warrant a simulation of the subsequent dynamics in that state. Also
notice that the energy $H_{0 0}$ of the extra orbital is irrelevant (so
one can take $H_{0 0}=0$) and that a single extra orbital is
sufficient regardless of the size of the system.

After each wavefunction collapse, the use of (\ref{classical}) implies
that the nuclei are treated classically. It is then appropriate to use
the mixed classical-quantum action~\cite{allen_1994,review} 
$S=\int dt\,L$, where
\begin{eqnarray}
L=\frac{1}{2}\langle \Psi _{e}|\left( i\hbar \frac{\partial }{\partial t}-
{\cal H}_{e}\right) |\Psi _{e}\rangle
+h.c. \nonumber \\
+\frac{1}{2}\sum_{k \alpha }\,
M_{k}\left( \frac{dX_{k \alpha }}{dt}\right) ^{2}
-U_{rep} 
\end{eqnarray}
where ${\cal H}_{e}$ is the electronic Hamiltonian, $|\Psi
_{e}\rangle$ is the electronic state, ``h.c.'' means ``Hermitian
conjugate'', $k$ labels a nucleus with spatial coordinates $\alpha$, 
and $U_{rep}$ is the
repulsive interaction between nuclei or ion cores. As shown in 
Ref.~\cite{review} (but with $\boldsymbol{H}$ now given by 
(\ref{new_full_H})), extremalization of this action leads to 
(\ref{matrix_SE}) and
\begin{eqnarray}
M \frac{d^{2} X}{dt^{2}} = 
-\frac{1}{2}\sum_{n} \boldsymbol{\psi }_{n}^{\dagger } 
\cdot \left( \frac{\partial \boldsymbol{H}}{\partial X } 
 - i\hbar \frac{\partial \boldsymbol{S}}{\partial X}
\frac{\partial }{\partial t}\right) \cdot \boldsymbol{\psi}_{n} \nonumber \\
 + h.c. - \frac{\partial U_{rep}}{\partial X} 
\end{eqnarray}
if one makes the usual time-dependent 
effective-field approximation, with exchange and 
correlation represented by an effective one-electron potential, and the
electronic state represented by a single antisymmetrized product
wavefunction $\Psi_{e} \left( t \right)$.
Here $X$ is any nuclear coordinate and $M$ is the corresponding mass.  

The reduced Ehrenfest method described above combines the advantages
of Born-Oppenheimer simulations, which are valid when (\ref{BO_approx}) 
holds, and Ehrenfest simulations, which are suitable for treating the
vibronic transitions when (\ref{close}) holds, as the results of 
Refs.~\cite{torralva,dou,dou_2004,sauer_2007,sauer} 
have clearly demonstrated. The use of 
reduced Ehrenfest simulations should solve various problems that are
encountered in standard Ehrenfest simulations -- for example, the
apparent failure of TDDFT 
to correctly describe the isomerization of
retinal~\cite{torralva_2005}. 
One problem with TDDFT is that the energies of excited states are
not accurately described, but a potentially more severe problem in the case of
molecules is that TDDFT is a special case of standard Ehrenfest
dynamics, and as a result fails to yield a complete return to the
ground state following de-excitation near a conical intersection.
In a reduced Ehrenfest simulation, on the other hand, one correctly follows 
the nuclear dynamics for that fraction of the population of molecules 
which does return to the ground state, and which therefore should isomerize
more readily. Reduced Ehrenfest simulations are practical for large
molecules, and are still consistent with the true meaning of quantum amplitudes, 
which yield probabilities for the various outcomes that are observed at the 
classical level. 

Finally, it may be worth noting that the above treatment can be
straightforwardly generalized to other particles, relativistic
systems, and nonabelian gauge fields, with $\psi$ in (\ref{SE}) 
interpreted as a multicomponent field and the Hamiltonian of (\ref{hamil}) 
appropriately changed. It can also be used with 
many-body effects included through self-energy terms, in the
Kadanoff-Baym/Keldysh equations for time-dependent and nonequilibrium 
problems~\cite{bonitz,vanLeeuwen}. The chief limitation is the use of localized
basis functions and the approximation (\ref{approx}).

\section*{Acknowledgements}
This work was supported by the Robert A. Welch Foundation (Grant A-0929).
I also thank Yusheng Dou and Meng Gao for their helpful comments.

\end{document}